# Reliability-based Sizing of Energy Storage for Large Load Integration

**A. BERA[1], C. CHAUDHARY[2], D. PANDIT[1], C. NEWLUN[1], T. NGUYEN[1], R. BYRNE[1], J. MITRA[2]**
**[1]Sandia National Laboratories, Albuquerque, NM 87123**
**[2]Michigan State University, East Lansing, MI 48824**
**United States of America**

## SUMMARY

The number of large industrial and commercial loads connected to the bulk power system has been increasing rapidly worldwide. Emerging high-demand facilities, particularly data centers supporting artificial intelligence (AI) applications and cryptocurrency mining, are introducing new challenges for power system planners as electricity demand continues to grow. In absence of appropriate planning mechanisms, the addition of high volumes of data center loads will lead to an increase in the magnitude, duration, and frequency of outages, thus degrading the reliability of the system. Energy storage systems (ESS) offer a promising solution to mitigate the reliability impacts of these large loads, owing to their operational flexibility and fast-ramping characteristics. However, ESS should be judiciously sized and sited to maximize the reliability benefits to the grid. This paper presents a framework for the optimal sizing and siting of ESS to minimize the adverse reliability impacts of large load additions. First, a stochastic semi-Markov model is developed to represent the load behavior of AI data centers and generate load profiles for several such facilities. These profiles are then incorporated into a Markov-chain Monte Carlo simulation-based composite system reliability assessment framework. A detailed nodal network model is used to quantify the reliability impacts of adding data centers and to identify the buses most affected by these large loads. Resource adequacy metrics are used to evaluate the overall system reliability impacts as well. ESS are then strategically sized and sited based on this analysis to mitigate the negative impacts of large loads on overall system reliability. The proposed framework is demonstrated using the RTS-GMLC 72-bus test system. The results show how ESS can effectively enhance system reliability in the presence of rapidly growing large-load demand. This framework provides system planners, utilities, regulators, and other stakeholders with a practical tool for maintaining reliable power system operation using energy storage as increasingly large and concentrated loads are integrated into the grid.

abera@sandia.gov

## 1. INTRODUCTION

AI-driven data center growth has outpaced the demand patterns conventional resource adequacy tools were built to accommodate. Individual facilities now draw hundreds of megawatts to 1 GW, and GPU cluster workloads impose sub-second power variability that static or slowly varying load models cannot capture [1], [2]. The North American Electric Reliability Corporation (NERC) has identified large electric loads as an emerging risk to bulk power system reliability [3] and has subsequently published reliability guidelines and mitigations [4]. The efforts by NERC highlight the need for robust planning methodologies capable of effectively evaluating and integrating large loads into future power systems. The mismatch between load behavior and planning assumptions motivates a reliability-based approach to sizing and siting energy storage for large load integration. Battery energy storage systems (BESS) can easily absorb this variability owing to their fast ramp rates and operational flexibility, but their reliability benefit depends on their size and location in the network.

Traditionally, BESS has been used in data centers primarily for uninterruptible power supply (UPS) applications [5]. However, as data centers pose increasing challenges to power systems, more advanced roles for BESS are being investigated. One important application is the use of on-site BESS to smooth data center load profiles. Recent work in [6] demonstrates the ability of grid-forming storage to operate as UPS while simultaneously smoothing AI-driven load transients. Similarly, [2] shows that rapid charging and discharging capability, high ramp rates, and sufficient energy capacity make BESS a highly effective solution to stabilize AI workload power demand, outperforming software-only and GPU-level power-smoothing mitigation techniques. At the grid scale, [7] identifies data centers as potential sources of power-system flexibility, largely due to the capabilities of storage systems within these facilities. Reference [8] further examines the potential roles of BESS in mitigating grid-reliability concerns associated with data center operations. Although BESS can provide multiple services that improve data center demand profiles and support grid reliability, appropriate sizing is essential to fully realize these benefits in an economical manner.

Existing research has examined data center and BESS sizing with different objectives. References [9], [10] propose BESS sizing approaches primarily aimed at ensuring sufficient backup generation for data centers. Similarly, [11] presents an optimal sizing framework for a data center and its co-located BESS that balances quality-of-service requirements with operational costs. Other studies expand the scope of their BESS sizing algorithms beyond data center facility-scale backup power. For example, [12] considers both backup power provision for the data center and frequency containment reserve services for the grid, while accounting for investment and operating costs, to determine the optimal BESS size. More recently, [13] proposes a BESS sizing algorithm that considers both data center demand-supply reliability and the provision of grid flexibility services. Although these studies address data center operational reliability and the use of BESS for grid services, they do not explicitly develop a grid reliability-based sizing strategy for BESS. Reliability-based sizing is important because transmission constraints, network topology, and outage scenarios can significantly affect the optimal placement or capacity of storage. As a result, sizing a BESS solely from the perspective of the data center facility, as in prior works, may not lead to a system-wide reliable solution.

This paper aims at filling this research gap by developing a reliability-based planning framework for sizing and siting of ESS to mitigate the adverse effects of rapid data center load addition to the bulk power system. A Markov-chain Monte Carlo simulation (MCS)-based composite system reliability assessment framework is developed to evaluate the reliability impacts of data center load addition. The data center load is represented using a stochastic semi-Markov model and synthetic load profiles thus generated are utilized by the MCS-based framework. A detailed reliability assessment is performed to evaluate the system reliability indices and identify the buses worst affected by data center load addition. ESS are then sized and sited based on this analysis to enhance system reliability. The main contributions of this paper are listed as follows.

1. A stochastic semi-Markov model is developed to represent the load behavior of AI data centers. This model considers the different subsystems of data centers with different operational timescales and dynamic behaviors. The proposed model is ideal for integration with stochastic reliability assessment frameworks as state transitions follow similar patterns in such processes.
2. A detailed composite system reliability assessment is performed using an MCS-based framework with nodal network model. This analysis not only yields the frequency-, duration-, and magnitude-based reliability metrics of the system, but also identifies the locations most adversely affected by data center load addition.
3. A strategy for sizing and siting ESS is proposed based on the reliability analysis. This strategy, along with the proposed reliability assessment framework, can be utilized by stakeholders including system planners, utilities, regulators, and researchers to optimally size and site ESS to maintain system reliability at desired levels as increasing volumes of data center loads are integrated into the bulk power system.

The remainder of this paper is organized as follows. Section 2 develops the stochastic semi-Markov load model for AI data centers. Section 3 presents the Monte Carlo simulation-based composite reliability evaluation framework. Section 4 describes the test system, the case studies, and the resulting reliability metrics. Section 5 concludes the paper and outlines directions for future work.

## 2. DATA CENTER LOAD MODEL

This section presents the stochastic load model for an AI data center, which generates the hourly load profiles that feed into the Monte Carlo reliability assessment. The model separates facility power into a temperature-dependent quasi-static base load and a stochastic IT compute load governed by a five-state semi-Markov process.

### 2.1 Load Decomposition

An AI data center comprises physically distinct subsystems with different timescales and dynamic behaviors. The aggregate active power $P_i(t)$ at the grid connection bus of facility $i$ at timestep $t$ decomposes as [14]

$$P_i(t) = P_{\text{base},i}(t) + P_{\text{IT},i}(t) \tag{1}$$

where $P_{\text{base},i}(t)$ is the quasi-static non-IT base load covering HVAC, chiller, and auxiliary systems, constant within each hour but varying seasonally with ambient temperature, and $P_{\text{IT},i}(t)$ is the stochastic IT load driven by GPU computation. $P_{\text{IT},i}$ is the sole source of rapid power variability. Component fractions, power factors, and timescales appear in Table 1. Fraction denotes each subsystem's share of the facility rated power $\bar{P}_i^{\text{rated}}$, and PF denotes the power factor of that subsystem.

*Table 1: Data center load components, power fractions, and characteristic timescales [2], [14].*

| Component | Fraction | PF | Timescale | Treatment |
|---|---|---|---|---|
| IT (GPU/CPU racks) | 60% | 0.995 | 0.1–30 s | Semi-Markov + OU |
| HVAC air-handling | 15% | 0.940 | $\tau = 60$ s | Quasi-static base |
| Chiller plant | 20% | 0.950 | $\tau = 120$ s | Quasi-static base |
| Auxiliaries | 5% | 0.970 | Quasi-static | Quasi-static base |

The base load scales with nameplate rating $\bar{P}_i^{\text{rated}}$ and ambient temperature [1], [3]:

$$P_{\text{base},i}(t) = [\alpha_{\text{cool}}(T_{\text{amb}}(h)) + \alpha_{\text{Aux}}]\,\bar{P}_i^{\text{rated}} \tag{2}$$

where $h \in \{1, \dots, 8760\}$ is the hour index and $\alpha_{\text{Aux}} = 0.05$. The cooling fraction $\alpha_{\text{cool}}$ scales linearly with ambient temperature relative to a reference of 15 °C, with nominal HVAC and chiller fractions of 0.15 and 0.20. Ambient temperature follows a sinusoidal annual profile with mean 12 °C, seasonal amplitude 12 °C, and peak in early July. These parameters take generic, literature-representative values rather than site-specific ones. Together, these parameterizations yield $P_{\text{base},i} \in [0.291,\ 0.466]\,\bar{P}_i^{\text{rated}}$ across the year. The base load therefore peaks in summer, when generation margins are tightest.

## 2.2 Five-State Semi-Markov IT Process

The IT component power $P_{\text{IT},i}(t)$ moves between a small number of operational regimes rather than varying continuously, and each regime holds for a duration set by the hardware itself. These hardware-imposed bounds rule out the memoryless exponential sojourn times of a standard continuous-time Markov chain. $P_{\text{IT},i}(t)$, therefore, follows a semi-Markov process on the state space $\mathcal{S} = \{\text{PS}, \text{LC}, \text{BA}, \text{CI}, \text{AR}\}$, calibrated to published power telemetry from DGX-H100 and H100/H200 clusters, Nvidia GPU hardware used for large-scale AI training [2], [15]. Each state holds for a random sojourn time, drawn from a uniform distribution over a finite, hardware-set interval $[T_s^{\min}, T_s^{\max}]$, before the system switches to the next state [16]. Table 2 lists the five states with their sojourn-time bounds and power levels. Here, AllReduce is a collective step that performs exchange of results between processors.

*Table 2: IT semi-Markov state parameterization [2], [15].*

| State | Full Name | Description | IT Power | $T^{\min}$–$T^{\max}$ (s) | sgn | $\delta$ range | $\sigma_{\text{IT}}$ |
|---|---|---|---|---|---|---|---|
| PS | Peak-Surge | Forward/backward propagation at thermal design power | 1.25–1.40 $\bar{P}_{\text{IT}}$ | 5–25 | +1 | U(0.25, 0.40) | 3.5% |
| LC | Light-Compute | Pipeline bubbles and gradient accumulation | 1.05–1.15 $\bar{P}_{\text{IT}}$ | 3–12 | +1 | U(0.05, 0.15) | 2.5% |
| BA | Base | Inter-iteration gap, kernel dispatch and memory management | 1.00 $\bar{P}_{\text{IT}}$ (ref) | 2–8 | 0 | 0 | 1.0% |
| CI | Checkpoint-Idle | Model checkpoint write to persistent storage | 0.75–0.85 $\bar{P}_{\text{IT}}$ | 30–180 | −1 | U(0.15, 0.25) | 1.5% |
| AR | AllReduce-Dip | Cross-node gradient synchronization via NCCL AllReduce | 0.50–0.65 $\bar{P}_{\text{IT}}$ | 2–15 | −1 | U(0.35, 0.50) | 2.0% |

The nominal hourly IT load for facility $i$ is

$$\bar{P}_{\text{IT},i}(h) = \alpha_{\text{IT}}\,P_i^{\text{rated}}\,\kappa_i^{(m)}\,\mu_D(h) \tag{3}$$

where $\alpha_{\text{IT}} = 0.60$; $\kappa_i^{(m)} \in [0.92, 1.08]$ is a monthly operating multiplier obtained from year-to-year load-factor variability documented in utility telemetry [1], [3]. $\mu_D(h) \in [0.55, 1.00]$ represents the daily variation in IT utilization. The daily utilization profile is modeled as a repeating 24-hour curve using six cosine-ramp segments. It assumes a minimum utilization of 55% from 00:00 to 06:00, a peak utilization of 100% from 10:00 to 18:00, and an evening utilization of 72% from 20:00 to 22:00, with smooth transitions between these periods. The profile is intended as a representative approximation rather than measurements from a specific facility. It is designed to reflect reported AI workload characteristics, including long periods of high utilization for model training and daily fluctuations in inference demand driven by user activity [3], [17], [18]. The power target in state $s$ is

$$P_{\mathrm{IT},i}^{\mathrm{tgt}}(t) = \bar{P}_{\mathrm{IT},i}(h)(1 + \mathrm{sgn}(s)\,\delta_s) \tag{4}$$

where $\delta_s \sim \mathrm{U}(\delta_s^{\min}, \delta_s^{\max})$ is a state-dependent deviation drawn at each state entry and $\mathrm{sgn}(s) \in \{+1, 0, -1\}$ for above-nominal, reference, and below-nominal states. An Ornstein-Uhlenbeck process, a random signal that wanders around a fixed level and relaxes back toward that level over time, is added on top of this target; this reproduces the small, sub-megawatt power fluctuations seen in measured GPU cluster traces [19]:

$$d\xi_{\mathrm{IT},i} = -\theta_{\mathrm{IT}}\,\xi_{\mathrm{IT},i}\,dt + \sigma_{\mathrm{IT}}(s_i)\,\bar{P}_{\mathrm{IT},i}(h)\,dW_{i,t} \tag{5}$$

where $dW_{i,t}$ is the Wiener increment, a random noise term, for facility $i$, $\sigma_{\mathrm{IT}}(s_i)$ is the state-dependent noise fraction, and $\theta_{\mathrm{IT}} = 0.8\,\mathrm{s}^{-1}$ gives a mean-reversion time of 1.25 s. The realized IT power is

$$P_{\mathrm{IT},i}(t) = \bar{P}_{\mathrm{IT},i}(h)(1 \pm \delta_s) + \xi_{\mathrm{IT},i}(t) \tag{6}$$

Transitions between state targets follow a linear ramp over 0.5 s.

The transition matrix for the different operational states is presented as follows [2], [15]:

$$\boldsymbol{A}_{(\mathrm{PS,LC,BA,CI,AR})} = \begin{bmatrix} 0 & 0 & 0 & 0 & 1 \\ 1 & 0 & 0 & 0 & 1 \\ 0 & 1 & 0 & 0 & 1 \\ 0 & 1 & 1 & 0 & 0 \\ 0 & 0 & 1 & 1 & 0 \end{bmatrix} \tag{7}$$

where $a_{ss'} = 1$ if the transition $s \to s'$ is physically possible and $p_{ss'} = 0$ whenever $a_{ss'} = 0$. PS exits only to AR. AR exits to CI or BA. CI exits to LC or BA. BA exits to LC or AR. LC exits to PS or AR. The canonical cycle is PS → AR → CI → LC → PS.

## 3. RELIABILITY EVALUATION FRAMEWORK

A sequential Monte Carlo simulation (MCS)-based composite system reliability analysis framework is used in this work. The composite system reliability analysis framework evaluates bulk power system adequacy by modeling the joint behavior of generation and transmission. The sequential MCS preserves the temporal dependencies of the different system components. The power system is represented by its primary components, including conventional generating units, variable energy resources (VERs), ESS, transmission lines, and loads. Conventional generators and transmission lines are represented using two-state Markov models with operational (*up*) and failed (*down*) states. VERs and ESS are represented using multi-state models, the details of which can be found in [20]. The MCS procedure employed for reliability assessment is described in detail in the following sections.

### 3.1 Mixed-time Sequential MCS

The sequential MCS framework utilized in this work for system reliability assessment is described in Algorithm 1.

**Algorithm 1. Sequential Monte Carlo Simulation for Reliability Assessment**

---

1: Read data including network topology, generator, ESS, load, data center load, component reliability, and meteorological data.
2: Initialize all generators and transmission lines to the *available* state.
3: Set sample path index: $s \leftarrow 1$.
4: **while** $s \leq N_{\text{samples}}$ **or** convergence criterion not satisfied **do**
  5: Set hourly index: $t \leftarrow 1$.
  6: **for** each generator and transmission line $i$ **do**
    7: Generate uniformly distributed variable: $U_i \sim \mathcal{U}(0,1)$.
    8: Compute initial transition time: $t_i = -\frac{1}{\lambda_i}\ln(U_i)$.
  9: **end for**
  10: Choose one set of data center demand profile randomly from user-provided data.
  11: **while** $t \leq 8760$ **do**
    12: Determine the next component transition time as: $t_{\min} = \min(t_1, t_2, \dots, t_{G+L})$.
    13: **if** $t_{\min} = 0$ **then**
      14: Toggle the corresponding component state.
      15: Regenerate its transition time.
    16: **end if**
    17: Decrease $t_{\min} \leftarrow t_{\min} - 1$.
    18: Determine the stochastic variable generation at all buses for hour $t$.
    19: Calculate net load as: $D_{b,t}^{\text{net}} = L_{b,t} - \sum_{v \in V_b} P_{v,t} + \sum_{z \in Z_b} D_{z,t}^{AI}$.
    20: **if** $t \bmod |T| = 0$ **then**
      21: Solve the multi-period dispatch problem as described in Section 3.2.
      22: **if** load curtailment occurs at any $\tau \in T$ **then**
        23: Record the event.
      24: **end if**
    25: **end if**
    26: $t \leftarrow t + 1$
  27: **end while**
  28: Using the recorded events for $s$, update the reliability indices.
  29: $s \leftarrow s + 1$
30: **end while**

---

**Notations:**
$\lambda_i$ is the failure or repair rate of component $i$,
$t_i$ is the time until the next state transition of component $i$,
$G$ and $L$ denote the sets of conventional generators and transmission lines, respectively,
$V$ and $Z$ denote the sets of variable generators and AI data centers, respectively,
$D_{b,t}^{\text{net}}$ is the net load at bus $b$,
$L_{b,t}$ is the demand at bus $b$,
$P_{v,t}$ is the variable generation at bus $b$,
$D_{z,t}^{AI}$ is the AI data center demand at bus $b$.

### 3.2 Optimal dispatch framework

In this work, multi-period economic dispatch problems are solved over the specified scheduling horizon using the available generation and net-load. This formulation captures the contribution of ESSs toward reducing system operating costs while maintaining system adequacy, and enables the quantification of the storage capacity required to achieve a target reliability level.

The objective function for a scheduling horizon $t \in T$ is defined as follows:

$$\sum_{t \in T} \left( \sum_{g \in G} C_g \, P_{g,t} + \sum_{s \in S} \left( C_s^{\mathrm{ch}} P_{s,t}^{\mathrm{ch}} + C_s^{\mathrm{dis}} P_{s,t}^{\mathrm{dis}} \right) + \sum_{b \in B} C_b^{\mathrm{curt}} \, P_{b,t}^{\mathrm{curt}} \right) \tag{8}$$

where $G$, $S$, and $B$ denote the sets of generators, ESSs, and system nodes, respectively. The parameters $C_g$, $C_s^{\mathrm{ch}}$, $C_s^{\mathrm{dis}}$, and $C_b^{\mathrm{curt}}$ represent the cost coefficients associated with generator output, ESS charging, ESS discharging, and load curtailment, respectively. The corresponding decision variables are $P_{g,t}$, $P_{s,t}^{\mathrm{ch}}$, $P_{s,t}^{\mathrm{dis}}$, and $P_{b,t}^{\mathrm{curt}}$. A sufficiently large penalty is assigned to load curtailment to encourage the utilization of all available resources before shedding load.

The optimization problem is subject to the following constraints:

1. *Power balance constraint:* This constraint ensures that the net power injection for all buses equal the net-load at that bus, and it is expressed as follows:

$$\sum_{l \in L} A_{b,l}^{\mathrm{inc}} \, F_{l,t} + \sum_{g \in G_b} P_{g,t} + \sum_{s \in S_b} P_{s,t}^{\mathrm{ESS}} + P_{b,t}^{\mathrm{curt}} = D_{b,t}^{\mathrm{net}} + P_{b,t}^{\mathrm{VG-curt}}, \qquad \forall b \in B, \forall t \in T \tag{9}$$

where $L$ is the set of transmission lines, $G_b$ and $S_b$ are the sets of generators and ESSs connected to bus $k$, respectively, $A^{\mathrm{inc}}$ is the bus-line incidence matrix, $F_{l,t}$ is the power flow for line $l$, $P_{s,t}^{\mathrm{ESS}}$ is the net ESS power injection, and $D_{b,t}$ is the load demand, and $P_{b,t}^{\mathrm{VG-curt}}$ is the variable generation curtailment.

2. *Generator capacity limits:*

$$P_{g,t}^{\min} \leq P_{g,t} \leq P_{g,t}^{\max}, \qquad \forall g \in G, \forall t \in T \tag{10}$$

3. *Transmission flow limits:*

$$F_l^{\min} \leq F_{l,t} \leq F_{l,t}^{\max}, \qquad \forall l \in L, \forall t \in T \tag{11}$$

4. *ESS state-of-charge constraint:* The ESS state of charge (SOC) constraint is enforced as follows:

$$s_{s,t} = s_{s,t-1} + \eta P_{s,t-1}^{\mathrm{ch}} - P_{s,t-1}^{\mathrm{dis}}, \tag{12}$$
$$s_s^{\min} \leq s_t \leq s_s^{\max}, \qquad \forall s \in S, \forall t \in T$$

where $s_t$ denotes the SOC at time $t$ and $\eta$ is the ESS charging efficiency.

5. *ESS charging/discharging limits:*

$$P_{s,t}^{\mathrm{ch}} + P_{s,t}^{\mathrm{dis}} \leq P_s^{\max} \qquad \forall s \in S, \forall t \in T \tag{13}$$

where $P_s^{\max}$ is the ESS power rating.

### 3.3 Data Center Load Integration

The AI data center load profiles developed in Section 2 are incorporated into the MCS. Although these profiles are available at a sub-hourly resolution, the MCS evaluates system states on an hourly basis. To reconcile this difference in temporal granularity, the peak data center load within each hour is used as the representative hourly value. This assumption is appropriate because the analysis is intended to capture a worst-case operating condition, consistent with the objective of sizing ESS to accommodate the additional data center demand.

To represent the inherent uncertainty in AI data center demand within the probabilistic resource adequacy assessment, multiple candidate load profiles are generated for each data center in the system. At the beginning of each MCS sample, one profile is randomly selected for each data center and used for that sample. This approach preserves the stochastic nature of the load modeling process and enables the assessment to capture a broader range of plausible operating conditions. As a result, the reliability evaluation reflects not only conventional system uncertainties but also the variability and uncertainty associated with data center demand.

### 3.4 Reliability Indices

The MCS framework produces outage statistics that are aggregated to calculate system reliability indices. The metrics used in this work to quantify system reliability are listed as follows:

1. *Loss of Load Probability (LOLP):* The probability that system demand exceeds the available supply capacity during a given hour.
2. *Loss of Load Expectation (LOLE) (days/year):* The expected number of days per year during which the available generation capacity is insufficient to meet the daily peak demand.
3. *Expected Unserved Energy (EUE) (MWh):* The expected amount of energy that cannot be supplied to customers over a specified time horizon. EUE may also be reported in normalized form (nEUE), expressed as a percentage of the total energy demand during the study period.
4. *Loss of Load Frequency (LOLF) (outages/year)*: The expected number of loss-of-load events occurring within a specified period, typically measured as the average number of occurrences per year.
5. *Mean Down Time (MDT) (h):* The average duration of load interruption events experienced during the simulation period.

## 4. CASE STUDIES, RESULTS, & ANALYSIS

This section provides a description of the test system, outlines the case studies, and presents the results accompanied by detailed analysis.

### 4.1 Test System

The RTS-GMLC [21] is used as a test system in this work. The test system, illustrated in Figure 1, consists of three balancing areas or zones, 72 buses, 120 lines, 93 conventional thermal and hydro power units, several VERs, and one ESS. A full nodal network model is utilized in this work, which considers all these components. The test system does not have access to external markets. The following modifications are made for the case studies:

1. The original test system is observed to be overbuilt, which leads to unrealistic values of reliability indices. Hence, the hourly load values are increased by 12% to obtain more realistic estimates of reliability indices.
2. The test system only provides data for one weather year for each VER site. However, data spanning multiple weather years are required for evaluating multiple scenarios using MCS. The required meteorological data, from the year 2000 to 2025 were obtained from the ERA5 reanalysis dataset through the Copernicus Climate Data Store (CDS) Application Programming Interface [22], [23].

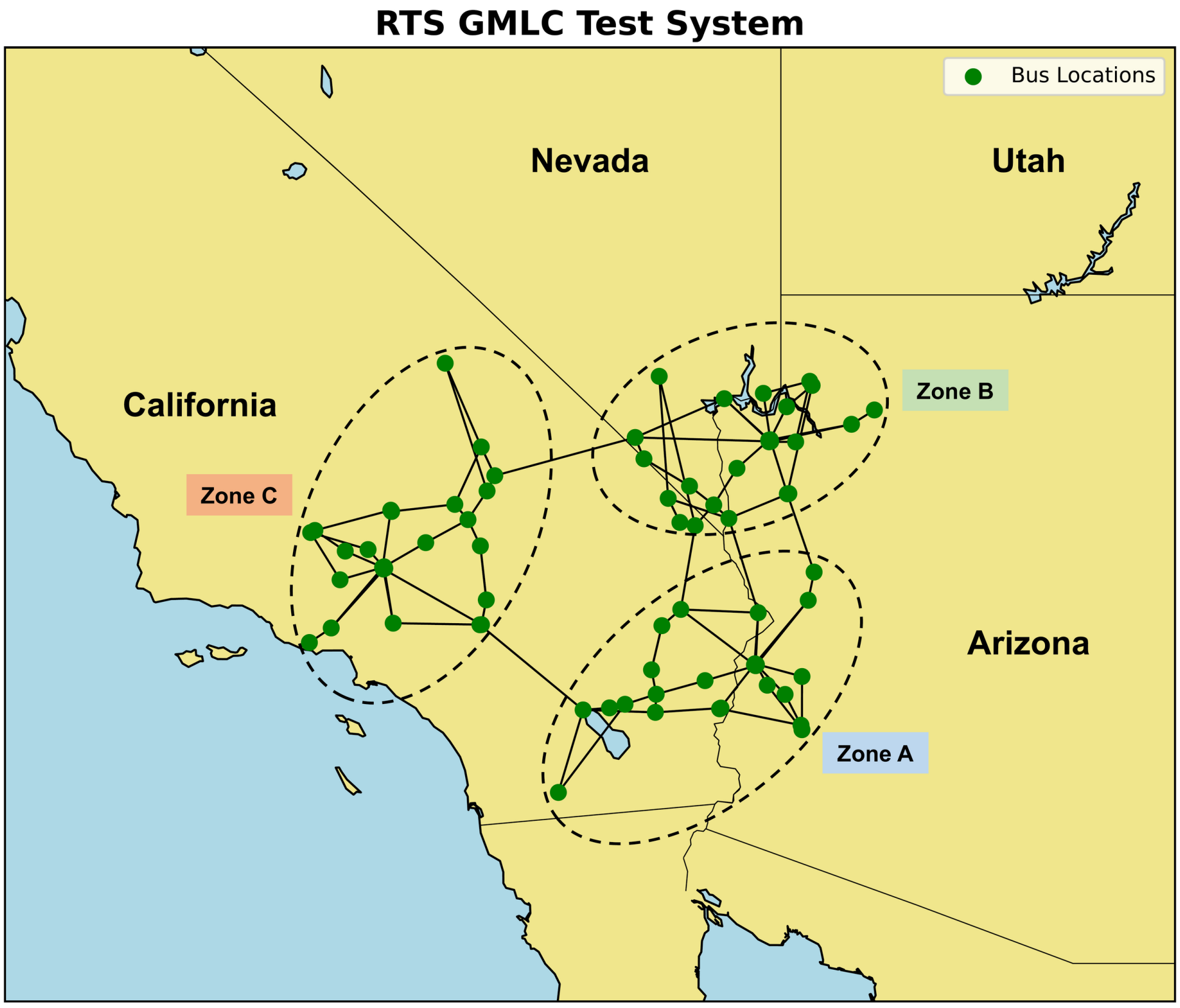

*Figure 1: Map of the RTS GMLC system*

## 4.2 Cases

Three case studies are performed in this work to demonstrate the reliability impacts of adding AI data center loads to the system and the role of ESS in mitigating these negative impacts. These cases are described as follows.

1. **Case I**: Reliability indices for the test system as described in the previous section are evaluated to establish a baseline.
2. **Case II**: For this case, three data center loads of sizes 200 MW, 230 MW, and 250 MW are added at buses Anna (Zone A), Bardeen (Zone B), and Cary (Zone C) respectively. These buses are chosen as they did not have any load or generation associated to them in the original test system and they have connections with multiple other buses. This strategy allows system operators to minimize load curtailment and transmission congestion events.
3. **Case III**: For this case, ESS are added to improve the reliability metrics of the data center-integrated system. Both 4-hr and 8-hr Lithium-ion batteries are considered as Cases III.a and III.b, respectively, to study how ESS duration affects reliability improvement. The strategy behind the sizing and siting of ESS is discussed in Section 4.3.

## 4.3 Results & Analysis

The system reliability metrics for the case studies are presented in Table 3. The metrics are obtained by running between 5,000–10,000 samples of the MCS (as described in Algorithm 1) for the different cases, depending on the convergence criteria.

*Table 3: System reliability metrics before and after adding data center loads.*

| Case | LOLP | LOLE (days/year) | EUE (MWh/year) | LOLF (outages/year) | MDT (h) |
|---|---|---|---|---|---|
| I | 0.0001 | 0.2595 | 38.3741 | 0.2571 | 0.8689 |
| II | 0.0011 | 2.5481 | 1839.9315 | 2.8009 | 2.7793 |

Results show that system LOLE increased almost ten-fold compared to the base case (Case I) after the addition of large loads in the system (Case II), while the EUE increased by almost 50 times. This demonstrates how addition of even modestly sized data centers can greatly affect system reliability. Further analysis shows how the individual buses are affected, with average outage seen by the three worst affected buses reported in Table 4.

*Table 4: Outage statistics and ESS added for buses most affected after data center load integration*

| Bus | Zone | Average Outage (MW/hr) | BESS Added (MW) |
|---|---|---|---|
| Baker | B | 112.77 | 115 |
| Carew | C | 107.81 | 110 |
| Adams | A | 52.2 | 60 |

A crucial observation from Table 4 is that although the data centers are added to buses Anna, Bardeen, and Cary as mentioned in Section 4.2, the buses worst affected by this load addition are completely different from these buses. Furthermore, in our detailed analysis, no outage was detected for buses Anna and Bardeen, while only a one-hour outage was detected for bus Cary for the duration of the simulation. This can be attributed to several factors including network topology, location of generating resources and loads in the system, and component reliability characteristics, which are system dependent. These results emphasize the importance of the proposed reliability-based planning framework and shows how adding data centers without co-located generation or ESS might impact the reliability of adjacent sections of the grid.

Finally, for Case III, 4-hour and 8-hour BESS are added to the three most affected buses. The power capacities of the BESS are decided based on the average outage magnitudes recorded at these buses, as shown in Table 4. These numbers are chosen for demonstration purposes. System planners may elect to install bigger or smaller BESS based on their reliability targets.

*Table 5: System reliability metrics after installing BESS.*

| Case | LOLP | LOLE (days/year) | EUE (MWh/year) | LOLF (outages/year) | MDT (h) |
|---|---|---|---|---|---|
| III.a | 0.0003 | 0.7310 | 699.0083 | 0.8290 | 1.5449 |
| III.b | 0.0003 | 0.6978 | 424.6405 | 0.7667 | 1.2549 |

The reliability metrics obtained for Case III are reported in Table 5. As expected, the system reliability metrics improve after batteries are installed, thus demonstrating the efficacy of the proposed framework. Comparing results between cases III.a and III.b, we observe that increasing BESS duration from 4 hours to 8 hours does not lead to a significant improvement in the duration and frequency indices (LOLP, LOLE, LOLF, MDT) but does reduce the magnitude of outages by almost 40%, as indicated by the EUE index. Further analysis is required to determine whether the investment cost justifies this improvement. This will be a part of our future studies.

## 5. CONCLUSION

This paper proposes a framework for the optimal sizing and siting of ESS to mitigate the adverse effects of large AI data center load additions on system reliability. A semi-Markov model is developed for representing AI data center loads. A Monte Carlo simulation-based composite system reliability analysis framework is used to determine the evolution of system reliability metrics as data center loads and ESS are added to the system. Results show that the buses at which data center loads are added are not necessarily the same buses that experience increased outages or degraded reliability. Instead, the affected buses may be located elsewhere in the system, depending on factors such as network topology and the spatial distribution of generation and load. While the installation of batteries leads to clear reliability benefits, increasing the BESS duration from 4 to 8 hours provides only limited improvement in duration- and frequency-based indices, suggesting diminishing returns in terms of outage frequency and duration. However, the substantial reduction in EUE indicates that longer-duration storage can still play an important role in mitigating the severity of outages by reducing the amount of unserved energy. These findings suggest that shorter-duration BESS may be sufficient to capture much of the reliability benefit, whereas longer-duration systems may still offer additional value in reducing outage magnitude. A comprehensive techno-economic assessment, via capacity expansion planning and production cost models, is needed in future work to determine whether the added investment required for longer-duration BESS is justified by the incremental reliability gains.

## ACKNOWLEDGMENT

This material is based upon work supported by the U.S. Department of Energy, Office of Electricity (OE), Energy Storage Division. This article has been authored by employees of National Technology & Engineering Solutions of Sandia, LLC under Contract No. DE-NA0003525 with the U.S. Department of Energy (DOE). The employees own all right, title and interest in and to the article and is solely responsible for its contents. The United States Government retains and the publisher, by accepting the article for publication, acknowledges that the United States Government retains a non-exclusive, paid-up, irrevocable, world-wide license to publish or reproduce the published form of this article or allow others to do so, for United States Government purposes. The DOE will provide public access to these results of federally sponsored research in accordance with the DOE Public Access Plan https://www.energy.gov/downloads/doe-public-access-plan.